\documentclass[journal=jacsat,manuscript=article]{achemso}

\usepackage[version=3]{mhchem} 
\usepackage{graphicx}
\usepackage{subcaption}
\usepackage{xcolor}

\author{Marco Antonio Barroca}
\affiliation{IBM Research, Rio de Janeiro, 20031-170, RJ, Brazil}
\altaffiliation{Centro Brasileiro de Pesquisas F\'isicas, Rio de Janeiro, 22290-180, RJ, Brazil}
\author{Rodrigo Neumann Barros Ferreira}
\affiliation{IBM Research, Rio de Janeiro, 20031-170, RJ, Brazil}
\email{rneumann@br.ibm.com}
\author{Mathias Steiner}
\affiliation{IBM Research, Rio de Janeiro, 20031-170, RJ, Brazil}

\title{Exploration of Quantum Computing in Materials Discovery for Direct Air Capture Applications}

\begin{document}



\begin{abstract}
  Direct air capture (DAC) of carbon dioxide is a promising method for mitigating climate change. Solid sorbents, such as metal-organic frameworks, are currently being tested for DAC application. However, their potential for deployment at scale has not been fully realized. The computational discovery of solid sorbents is challenging, given the vast chemical search space and the DAC requirements for molecular selectivity. Quantum computing can potentially accelerate the discovery of solid sorbents for DAC by predicting  molecular binding energies. In this work, we explore simulation methods and algorithms for predicting gas adsorption in metal-organic frameworks using a quantum computer. Specifically, we simulate the potential energy surfaces of CO\textsubscript{2}, N\textsubscript{2}, and H\textsubscript{2}O molecules at the Mg\textsuperscript{+2} metal center that represents the binding sites of typical metal-organic frameworks. We apply the qubit-ADAPT-VQE technique to run simulations on both classical computing and quantum computing hardware, and achieve reasonable accuracy while maintaining hardware efficiency.
\end{abstract}

\section{Introduction}

Technologies for capturing greenhouse gases (GHG) are essential climate mitigation tools that can help reduce GHG atmospheric concentration and limit global warming.\cite{Metz_2005} Existing energy generation infrastructure, such as natural gas and coal-fired power plants, can be retrofitted to perform post-combustion capture of carbon dioxide. This strategy can extend the lifetime of existing infrastructure and minimize the impact of energy generation to the global carbon budget. Nevertheless, merely bringing emissions to zero is not sufficient for achieving climate goals. Carbon dioxide removal from the atmosphere is needed for limiting the temperature increase~\cite{international2022direct}. Direct air capture (DAC) is an emerging carbon removal technology that employs sorbents to capture CO\textsubscript{2} directly from the atmosphere.

Liquid sorbent based technologies, such as amine scrubbing, are well-established for capturing carbon dioxide.\cite{sanz2016direct} Nevertheless, despite their fast adsorption rate and high CO\textsubscript{2} saturation, amines have certain disadvantages. They are corrosive, and due to their high desorption energy requirement, are responsible for 60-80\% of the total operating costs of CCS.\cite{Siegelman_2021, sabatino2021comparative} Solid sorbent materials for carbon dioxide capture offer an alternative approach to overcome the limitations of amine-based, liquid adsorption. This material class includes a diverse range of nanoporous solids, such as zeolites, metal-organic frameworks, covalent-organic frameworks, zeolitic imidazolate frameworks and porous polymer networks, among others. Porous solid sorbents can selectively adsorb CO\textsubscript{2} without bond formation (i.e., via physisorption), which drastically lowers the regeneration energy and, therefore, the operating costs.\cite{kumar2015direct} On the other hand, solid sorbents can be very sensitive to the presence of water vapor in the atmosphere, leading to decreased selectivity due to competitive adsorption with CO\textsubscript{2} and incomplete regeneration due to water passivation of adsorption sites.\cite{decoste2013effect}

Among the solid materials currently under investigation, Metal-Organic Frameworks (MOFs) have attracted attention due to their customizable porous structure and large chemical diversity.\cite{Sumida_2012} MOFs are composed of two types of building blocks, i.e., organic linkers and metal ions, which can be combined to form nanoporous structures with channels whose diameters range from a few to tens of Angstroms. Their chemical and geometrical aspects can be modified, to a large extent, independently for optimizing both the absolute (adsorption capacity) and relative (adsorption selectivity) metrics that determine capture and separation performance.\cite{yao2021inverse} The large variety of MOFs poses significant computational challenges for the selection of the best candidate material for a given application. There are databases of hypothetical MOFs,\cite{Wilmer_2012, Boyd_2019} as well as experimentally validated MOFs,\cite{Moghadam_2017} containing $10^{5} - 10^{6}$ materials in total.

\begin{figure}[ht]
  \centering

  \begin{subfigure}[b]{0.45\textwidth}
    \centering
    \includegraphics[width=\textwidth]{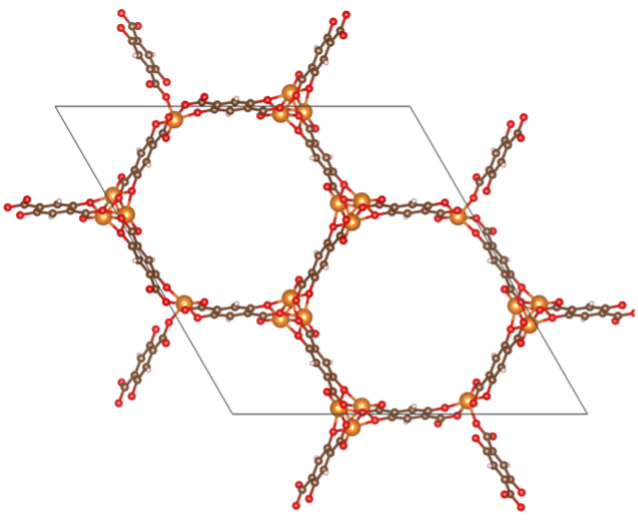}
    \caption{Conventional unit cell with 162 atoms}
    \label{fig:MOF_conv}
  \end{subfigure}
  \quad
  \begin{subfigure}[b]{0.45\textwidth}
    \centering
    \includegraphics[width=\textwidth]{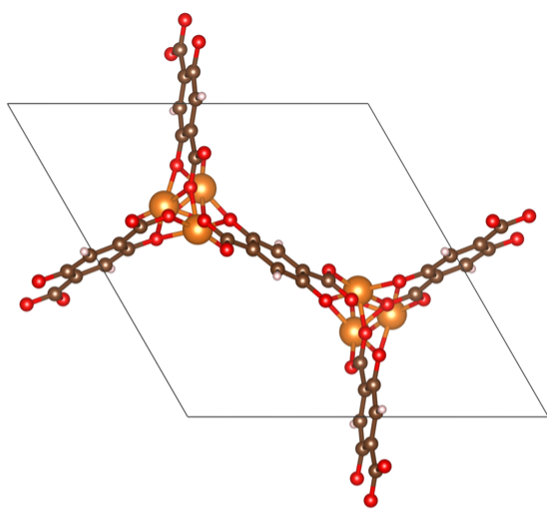}
    \caption{Primitive unit cell with 54 atoms.}
    \label{fig:MOF_primitive}
  \end{subfigure}

  \caption{Atomistic visualisation of Mg-MOF-74 structure}
  \label{fig:Mof_struc}
\end{figure}

In this work, we perform Quantum Chemistry calculations, on classical and on quantum computing hardware, in the form of the superconductor-based quantum computers available in the IBM Quantum platform, to estimate the binding potential of a set of representative molecules in the atmosphere (CO\textsubscript{2}, H\textsubscript{2}O and N\textsubscript{2}) within a simplified model system. We limited the search space for a carbon capture material by focusing on the X-MOF-74 family (where X=Mg\textsuperscript{+2}, Ni\textsuperscript{+2}, Fe\textsuperscript{+2}, Co\textsuperscript{+2}, Zn\textsuperscript{+2}, Mn\textsuperscript{+2}, Cu\textsuperscript{+2}) as a template for exploring the chemical diversity in MOFs. The MOF-74 family is known for the presence of coordinatively unsaturated metal ions (open metal sites) that can form a bond with adsorbate molecules, greatly influencing the adsorption strength and selectivity, especially at low pressures. The formation of chemical bonds with the adsorbate molecule is a phenomenon known as chemisorption and determines, to a great extent, the desorption energy cost.\cite{kumar2015direct,decoste2013effect}

\subsection{Methods}

\begin{figure}[ht]
  \centering

  \begin{subfigure}[b]{0.3\textwidth}
    \centering
    \includegraphics[width=\textwidth]{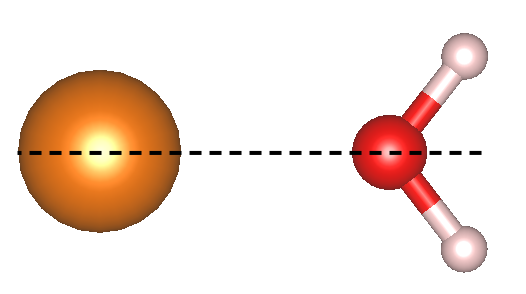}
    \caption{$M$ + H\textsubscript{2}O}
    \label{fig:Mol_H2O}
  \end{subfigure}
  \quad
  \begin{subfigure}[b]{0.3\textwidth}
    \centering
    \includegraphics[width=\textwidth]{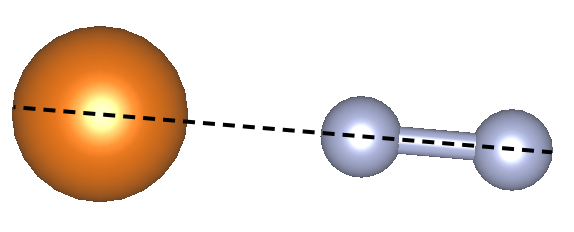}
    \caption{$M$ + N\textsubscript{2}}
    \label{fig:Mol_N2}
  \end{subfigure}
   \quad
  \begin{subfigure}[b]{0.3\textwidth}
    \centering
    \includegraphics[width=\textwidth]{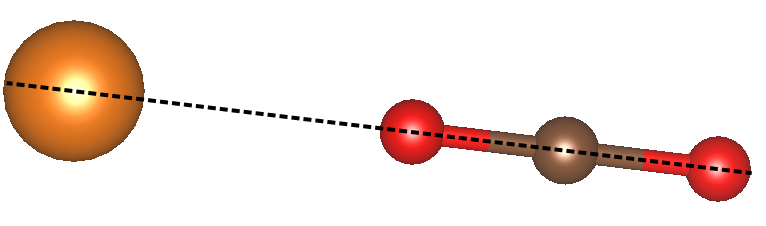}
    \caption{$M$ + CO\textsubscript{2}}
    \label{fig:Mol_CO2}
  \end{subfigure}
  \quad
\caption{Molecular geometries and bond axes used in the PES calculations for (a) H\textsubscript{2}O, (b) N\textsubscript{2} and (c) CO\textsubscript{2} molecules in the vicinity of a metal ion $M$. For H\textsubscript{2}O, the ion is moved along the C\textsubscript{2} rotation axis. For N\textsubscript{2}, the ion is moved along the N-N axis. For CO\textsubscript{2}, the ion is moved along the O-C-O axis}
\label{fig:molecules}
  
\end{figure}

Since the Mg-MOF-74 primitive cell, as shown in Figure \ref{fig:Mof_struc}, contains more than 50 atoms, we simplify the system even further by considering a single metal ion representing an open metal site and interacting with the adsorbate molecules. We do so because of the current limitations of quantum hardware which limits the width and depth of quantum circuits that can be reliably executed. We calculate Potential Energy Surfaces (PES) for several metal/molecule combinations and extract the equilibrium distance and binding energy for the systems represented in Figure \ref{fig:molecules}. This proof-of-concept illustrates how one can use quantum computing to address the challenge of discovering materials for direct air capture.

\subsubsection{Species selection}

Given the relevance of the metal ions to DAC applications of MOFs, we chose to focus on Mg\textsuperscript{+2}
metal ion and its pairings with all the three provided gas molecules (H\textsubscript{2}O, CO\textsubscript{2}, and N\textsubscript{2}), amounting to six distinct systems. In short, Mg\textsuperscript{+2} + H\textsubscript{2}O should be the simplest system as it contains the least number of electrons.

\subsubsection{Simulation methods}

We chose PySCF,\cite{Sun_2018, Sun_2020} as the driver for performing classical benchmarks. We performed two types of classical benchmarks with PySCF: Hartree-Fock (HF) and Density Functional Theory (DFT). The chosen basis set was 6-31g* and was kept consistent across all calculations, including for the quantum calculations. For DFT we chose the B3LYP functional. This particular choice of parameters was motivated by the methodology employed to compute 3D structures (SD-file) of the H\textsubscript{2}O, CO\textsubscript{2}, and N\textsubscript{2} molecules found in the NIST Chemistry WebBook, from which we also took the molecular structures used in the calculations. \cite{NIST_WebBook}

\subsubsection{Considerations for Simulations on Quantum Computers}

We chose Qiskit\cite{Qiskit} as the driver for performing quantum simulations using the Variational Quantum Eigensolver (VQE).\cite{Peruzzo_2014, Romero_2018, McClean_2016} We employed three methodological variants:

\begin{itemize}
    \item A standard VQE algorithm with an application-specific, compute-heavy UCCSD ansatz,\cite{Evangelista_2019} to be treated as the quantum benchmark, as it provides the best possible result for this kind of application. Unfortunately, it is challenging to use this approach on current quantum hardware, due to its complexity.
    \item A standard VQE algorithm with a general-purpose, hardware-efficient heuristic ansatz~\cite{tilly2022variational} that can run on current quantum hardware. In our case, we determined that an ansatz with R\textsubscript{Y} and R\textsubscript{Z} rotations, which we call R\textsubscript{Y}R\textsubscript{Z}, converges to the intended result despite requiring more iterations.
    \item A modified ADAPT-VQE algorithm\cite{Grimsley_2019} that utilizes qubit operators in the operator pool. This method is called qubit-ADAPT-VQE and has the advantage of giving similar results to UCCSD while having much lower requirements on the quantum hardware.\cite{Tang_2021}
\end{itemize}

\subsubsection{Active Space Selection}
To limit the computational cost for the VQE calculations, we performed a suitable active space selection. We used a known method implemented within PySCF called atomic valence active space (AVAS). AVAS is an automated method to construct an active space for electronic structure calculations in strongly correlated chemical systems, where single-determinant wave functions are inadequate. It focuses on identifying specific valence atomic orbitals, such as transition metal \textit{d} orbitals or those involved in bond dissociation, as a basis for defining the active space. The approach leverages density matrix embedding theory (DMET) to determine the minimal set of orbitals required to capture the essential electronic features, aiming for an active space no more than twice the size of the initially chosen orbitals. This method does not differentiate between ``fragment'' and ``bath'' orbitals, instead maintaining the occupied and virtual character of active orbitals, facilitating a natural truncation process to minimize active space size while capturing critical determinants.\cite{Sayfutyarova_2017}

\section{Results and discussion}

\subsection{Convergence studies}

\begin{figure}[ht]
  \centering
  \includegraphics[width=0.95\textwidth]{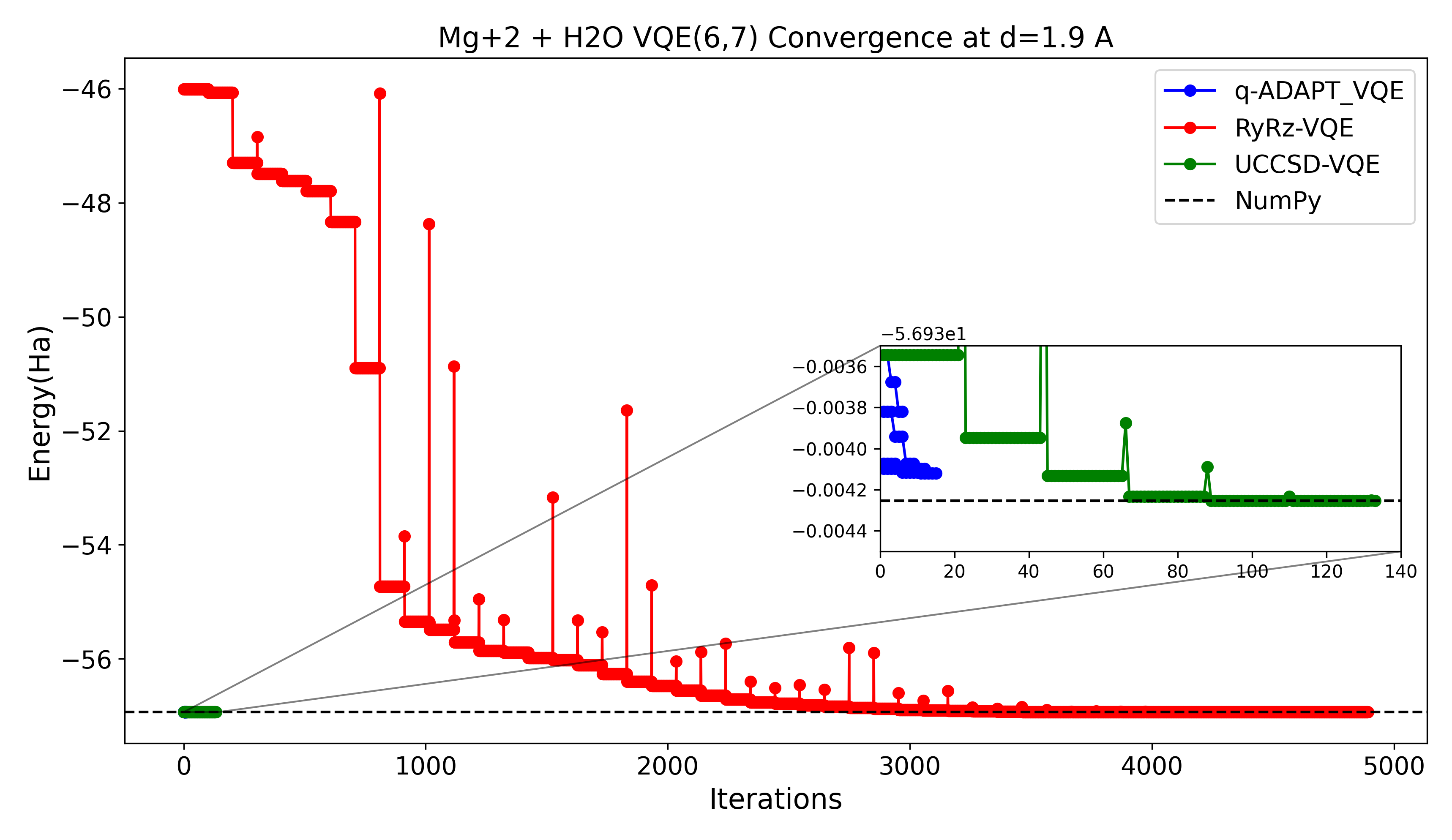}
  \caption{Convergence to the ground state energy calculated by UCCSD-VQE (green), R\textsubscript{Y}R\textsubscript{Z}-VQE (red) and qubit-ADAPT-VQE (blue) as a function of number of iterations. The result of the direct diagonalization of the fermionic Hamiltonian using NumPy (black) is shown as the target. All ansatze converge, to within chemical accuracy ($\sim 1\ \text{mHa}$), of the target result. However, the qubit-ADAPT-VQE ansatz reaches convergence in considerably fewer iterations than the UCCSD-VQE and R\textsubscript{Y}R\textsubscript{Z}-VQE ansatze. The energy values correspond to the computed part in the (6,7) active space.}
  \label{fig:Mg_qad}
\end{figure}

We begin by evaluating the performance of qubit-ADAPT-VQE against the more standard VQE approaches. In Figure \ref{fig:Mg_qad}, we compare all these VQE simulations in terms of their convergence and accuracy with respect to the ground state energy value obtained from the direct diagonalization of the Hamiltonian using NumPy. We used the Mg\textsuperscript{+2} + H\textsubscript{2}O model system at 1.9 \AA\ separation and an active space of six electrons and seven molecular orbitals (6,7). In Table \ref{tbl:runtime}, we display the runtime and total number of iterations needed for the convergence of each VQE ansatz. The qubit-ADAPT-VQE took 3x and 64x fewer iterations, respectively, than the UCCSD-VQE and R\textsubscript{Y}R\textsubscript{Z}-VQE ansatze. In terms of run time, the qubit-ADAPT-VQE was 5x and 12x faster than the UCCSD-VQE and R\textsubscript{Y}R\textsubscript{Z}-VQE ansatze, respectively.

\begin{table}
  \caption{The qubit-ADAPT-VQE ansatz converges in 3x and 116x fewer iterations, and is 5x and 25x faster than the UCCSD-VQE and R\textsubscript{Y}R\textsubscript{Z}-VQE ansatze, respectively.}
  \label{tbl:runtime}
  \begin{tabular}{llll}
    \hline
        & UCCSD-VQE & R\textsubscript{Y}R\textsubscript{Z}-VQE  & qubit-ADAPT-VQE  \\
    \hline
    Run time (s)  & 25 & 151   & 6  \\
    Total \# of iterations  & 133 & 4892   & 42  \\
    Energy Value (Ha)   & -56.93440792 & -56.93383614   & -56.93411960  \\
    \hline
  \end{tabular}
\end{table}

The qubit-ADAPT-VQE ansatz accomplishes fast convergence while remaining hardware efficient, as we can see from Table \ref{tbl:gates} comparing the sizes of the circuits between methods with the largest circuit that qubit-ADAPT-VQE uses. These circuit sizes are expressed in terms of the gates that IBM quantum hardware can execute and were built using the transpilation procedure from Qiskit. The most important metric here is the number of CNOT gates as those have error rates that are orders of magnitude higher than the single-qubit gates.

\begin{table}
  \caption{Comparing the circuit sizes generated by different ansatze. qubit-ADAPT-VQE manages to get satisfactory results while using the shallowest circuits. Circuits are described in terms of the number of gate operations and depth required to run on IBM quantum hardware.}
  \label{tbl:gates}
  \begin{tabular}{llll}
    \hline
        & UCCSD-VQE & R\textsubscript{Y}R\textsubscript{Z}-VQE  & qubit-ADAPT-VQE  \\
    \hline
    \# of CNOT gates  & 1610 & 27   & 20  \\
    \# of Rz gates  & 1191 & 125  & 37  \\
    \# of Sx gates   & 878 & 80   & 19 \\
    \# of X gates  & 44 & 5   & 1  \\
    Depth & 3255 & 35   & 30  \\
    \hline
  \end{tabular}
\end{table}

The performance advantage of qubit-ADAPT-VQE comes from the fact that the ansatz grows on demand from an operator pool as the convergence thresholds require. We created the operator pool by first building an UCSSD ansatz and then removing Pauli strings with an even number of Y operations.\cite{Tang_2021} This way we are left only with the ``odd'' Pauli Strings which are the ones relevant for the energy calculation. We chose the SLSQP optimizer to run the simulations. Finally, we defined the convergence thresholds: maximum number of iterations as $4$, maximum operator gradient as $10^{-4}$, and maximum change of the energy expectation value between consecutive iterations as $10^{-3}$. These convergence thresholds ensure chemical accuracy ($\sim 1\ \text{mHa}$) with respect to the UCCSD-VQE and NumPy results. These settings were used for all qubit-ADAPT-VQE simulations.

We did not include a noise model in the simulations, as we are mostly interested in comparing how our ansatze and methods perform against a benchmark provided by the direct diagonalization of the Hamiltonian with NumPy on a classical computer. Only after their convergence and accuracy are well established, we can repeat the best procedure on the quantum hardware.

We had to choose a suitable active space to run these quantum simulations in a reasonable time. Whenever possible, we utilized the AVAS automated selection method mentioned before, but if the given active space required more than 15 qubits, we would default that calculation to an active space of six electrons and seven molecular orbitals (6,7) chosen around the HOMO-LUMO level with orbitals ordered by energy contribution. This active space requires 10-11 qubits, which we adopted as a reasonable compromise between quality and runtime. All VQE and PySCF simulations were executed on a laptop (Apple MacBook Air M1 2020). We ran the simulations on a slim environment so that anyone with access to our code repository could easily reproduce our results.\cite{git_repo}

\subsection{Mg\textsuperscript{+2} systems}

Due to the excellent cost-benefit obtained from the qubit-ADAPT-VQE ansatz in the simulations, we selected this as our best candidate to run the calculations on IBM Quantum hardware. However, to run these calculations on the available quantum hardware, we had to further reduce our active space to three electrons and four molecular orbitals (3,4), which only required 4 qubits. Despite nominally supporting higher qubit counts, current hardware are not suitable for reliably running circuits of the required depth for this application due to a lack of fault tolerance. Again, the molecular orbitals were chosen around the HOMO-LUMO level and ordered by energy. We set the limit on the number of shots to 1024, which takes from five to eight hours depending on the separation distance.

The 4-qubit active space limitation led us to focus only on the Mg\textsuperscript{+2} systems for the VQE calculations on quantum hardware which we refer to as hardware experiments. This active space corresponds to $\approx 5\%$ of the energy contributions to the ground state, therefore we saw no point in simulating systems with a higher electron count as it would mean an even lower representation of the results obtained. We used Qiskit Runtime and \texttt{ibmq\_montreal} to run all our quantum hardware experiments, and we employed error mitigation techniques such as Twirled Readout Error Extinction (T-REx).\cite{Berg_2022}. 

The PES was built by computing the energy for different distances between metal ions and molecules along the bond axis. For all simulations and hardware experiments, we evaluate the energy at distances between 0.3 and 3.6 \AA, in increments of 0.3 \AA.

After we calculate the energies, we fit a Morse potential\cite{Morse_1929} to the data points to help us calculate precise equilibrium distances and molecular binding energies. This potential is a suitable inter-atomic model for the energy of diatomic molecules. In the PES plots, we offset the energies so that the asymptotic limit at infinite separation coincides with the zero-energy point to facilitate the extraction of the binding energy.

We simulated Mg\textsuperscript{+2} binding with all three gas molecules using the three VQE methods mentioned before: R\textsubscript{Y}R\textsubscript{Z}-VQE, UCCSD-VQE, and qubit-ADAPT-VQE, on a (6,7) active space. We evaluated their accuracy by comparing their PES to that obtained via direct diagonalization of the Hamiltonian with NumPy. 

\begin{figure}[ht]
  \centering
  \begin{subfigure}[b]{0.3\textwidth}
    \centering
    \includegraphics[width=\textwidth]{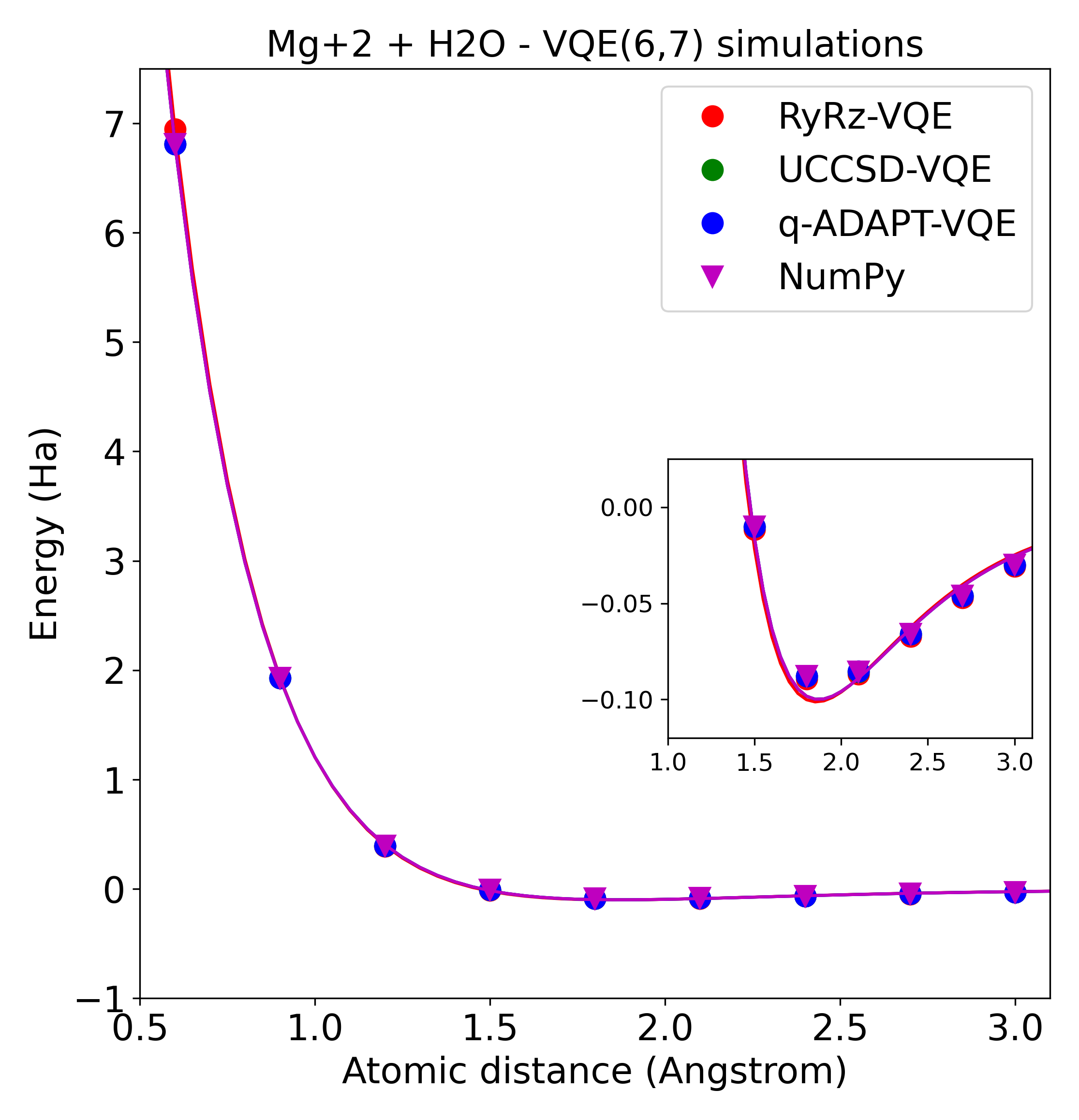}
    \caption{Mg\textsuperscript{+2} + H\textsubscript{2}O}
    \label{fig:Mg_H2O_VQE_plot}
  \end{subfigure}
  \quad
  \begin{subfigure}[b]{0.3\textwidth}
    \centering
    \includegraphics[width=\textwidth]{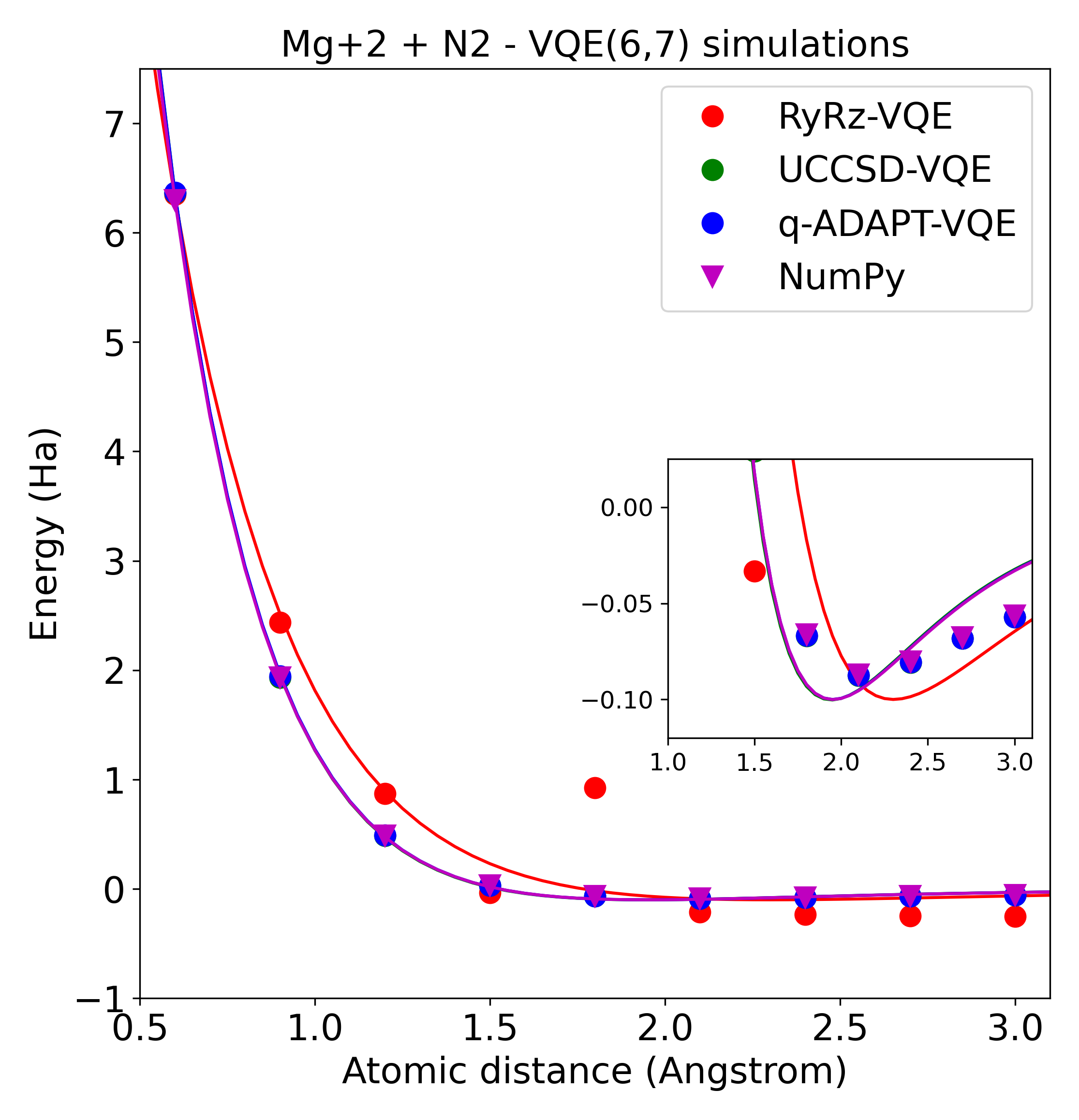}
    \caption{Mg\textsuperscript{+2} + N\textsubscript{2}}
    \label{fig:Mg_N2_VQE_plot}
  \end{subfigure}
   \quad
  \begin{subfigure}[b]{0.3\textwidth}
    \centering
    \includegraphics[width=\textwidth]{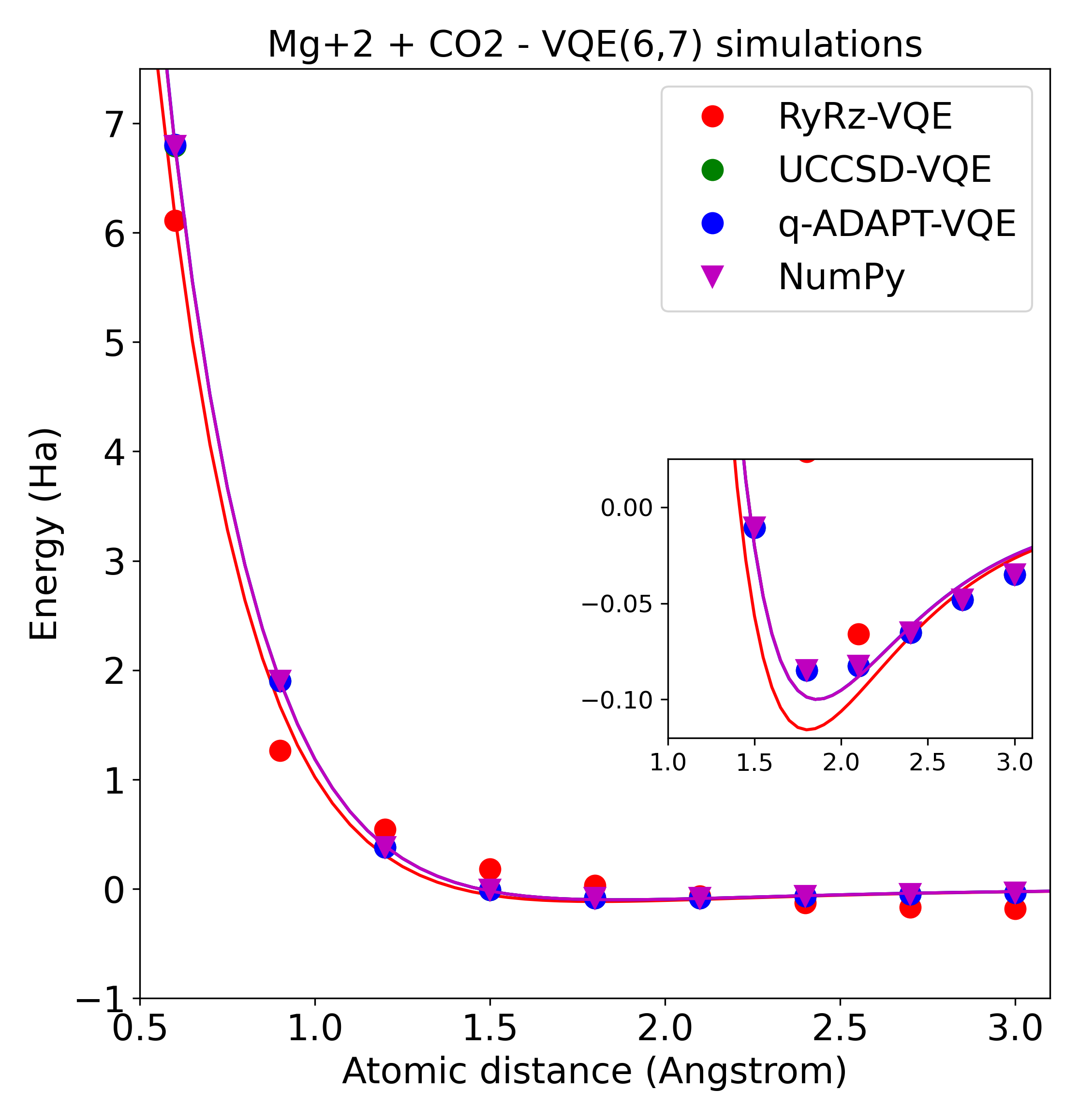}
    \caption{Mg\textsuperscript{+2} + CO\textsubscript{2}}
    \label{fig:Mg_CO2_VQE_plot}
  \end{subfigure}
  \quad
\caption{UCCSD-VQE (green circles) and qubit-ADAPT-VQE (blue circles) match the NumPy (magenta triangles) benchmark in all three cases, however, R\textsubscript{Y}R\textsubscript{Z}-VQE (red circles) only matches the benchmark for the simplest system, namely, Mg\textsuperscript{+2} + H\textsubscript{2}O.}
\label{fig:Mg_VQE_plots}
  
\end{figure}

In Figure \ref{fig:Mg_VQE_plots}, we observe that qubit-ADAPT-VQE and UCCSD-VQE successfully achieve the NumPy result, while R\textsubscript{Y}R\textsubscript{Z}-VQE only matches effectively the simplest system of the three, Mg\textsuperscript{+2} + H\textsubscript{2}O. The R\textsubscript{Y}R\textsubscript{Z}-VQE results for the N\textsubscript{2} and CO\textsubscript{2} systems overestimate the energy to the left of the minimum and underestimate to the right, affecting the Morse potential fit and leading to different observables.

Based on the accuracy of qubit-ADAPT-VQE results in Figure \ref{fig:Mg_VQE_plots}, and the superior performance over UCCSD-VQE demonstrated in Figure \ref{fig:Mg_qad}, Table \ref{tbl:runtime} and Table \ref{tbl:gates}, we confirm the choice of qubit-ADAPT-VQE for running VQE on quantum hardware. We limit ourselves to the smallest possible system, namely, Mg\textsuperscript{+2} + H\textsubscript{2}O for the reasons explained previously. We ran qubit-ADAPT-VQEwith an active space of (3,4) to comply with the limitations of current quantum hardware.

\begin{figure}[ht]
  \centering
  \includegraphics[width=0.5\textwidth]{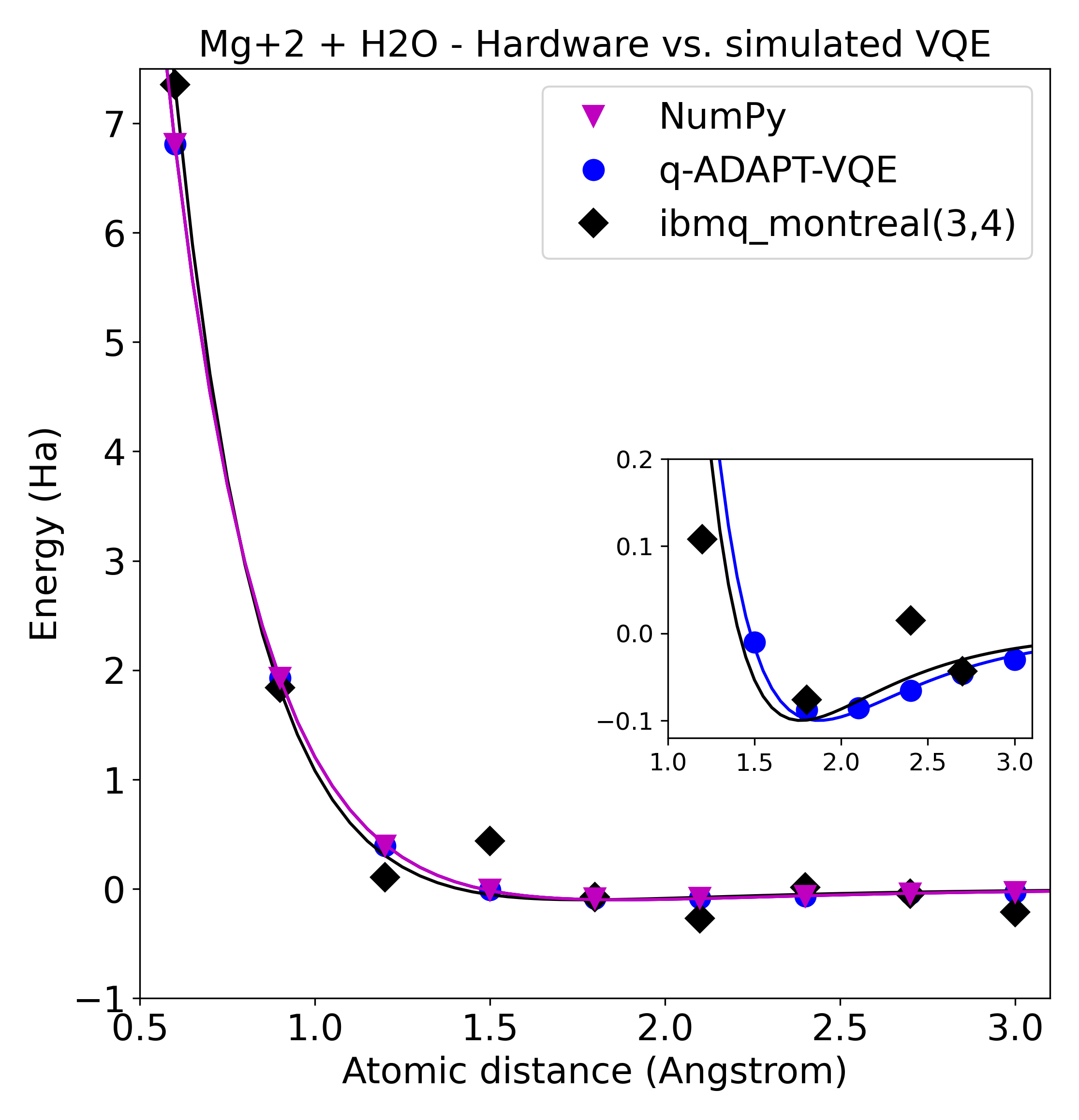}
  \caption{Potential Energy Surfaces obtained from running VQE on quantum hardware (black diamonds) compared to a qubit-ADAPT-VQE simulation (blue circles) and NumPy direct diagonalization (purple triangles). The adjusted Morse potentials exhibit a considerable overlap, despite the level of noise present in the hardware data points, which leads to a reasonable agreement between the observables shown in Table \ref{tbl:Mg_res}.}
  \label{fig:Mg_Rh_VQE}
\end{figure}

The results, shown in Figure \ref{fig:Mg_Rh_VQE}, also provide a comparison to a quantum hardware calculation. The Morse potential fit to the hardware result matches quite closely those of the direct diagonalization via NumPy and simulated VQE, despite the level of noise that is inherent to calculations on current quantum hardware. In this hardware experiment, we used 1024 shots and employed the T-REx noise mitigation technique which proved sufficient to handle the noise profile of the quantum device. Increasing the number of shots did not improve the accuracy of the experimental values, and using more robust error mitigation strategies, such as Zero Noise Extrapolation, would increase the already large variance in the results.

The plots show the calculated energies at selected distances represented by solid symbols and the Morse potential fit represented by solid lines. In the following plots, circles represent VQE simulations, triangles represent the direct diagonalization with NumPy, squares represent the PySCF calculations and diamonds represent the hardware experiments.

We extracted observables such as the equilibrium distance and the binding energy from the adjusted Morse potentials to the Mg\textsuperscript{+2} + H\textsubscript{2}O PES curves. Although we cannot claim to have achieved agreement to within chemical accuracy ($\sim 1\ \text{mHa}$), in Table \ref{tbl:Mg_res} we observe an excellent match between qubit-ADAPT-VQE and NumPy, and a reasonably good match with \texttt{ibmq\_montreal}, considering the higher uncertainties caused by noise in the quantum hardware.

\begin{table}
  \caption{Comparison of Equilibrium Distance and Binding Energy for Mg\textsuperscript{+2} + H\textsubscript{2}O obtained from the fitted Morse potential to results obtained with different methods. All methods are in agreement, to some extent, when we consider the uncertainties involved in the fitting procedure and the inherent noise of the quantum hardware data points.}
  \label{tbl:Mg_res}
  \begin{tabular}{llll}
    \hline
        & \texttt{ibmq\_montreal} & qubit-ADAPT-VQE & NumPy  \\
    \hline
    Equilibrium Distance (Å)  & $1.8\pm0.3$ & $1.87\pm0.01$   & $1.87\pm0.01$  \\
    Binding energy (Ha)  & $-0.1\pm0.4$ & $-0.1\pm0.02$  & $-0.1\pm0.02$  \\
    \hline
  \end{tabular}
\end{table}

\begin{figure}[ht]
  \centering
  \begin{subfigure}[b]{0.3\textwidth}
    \centering
    \includegraphics[width=\textwidth]{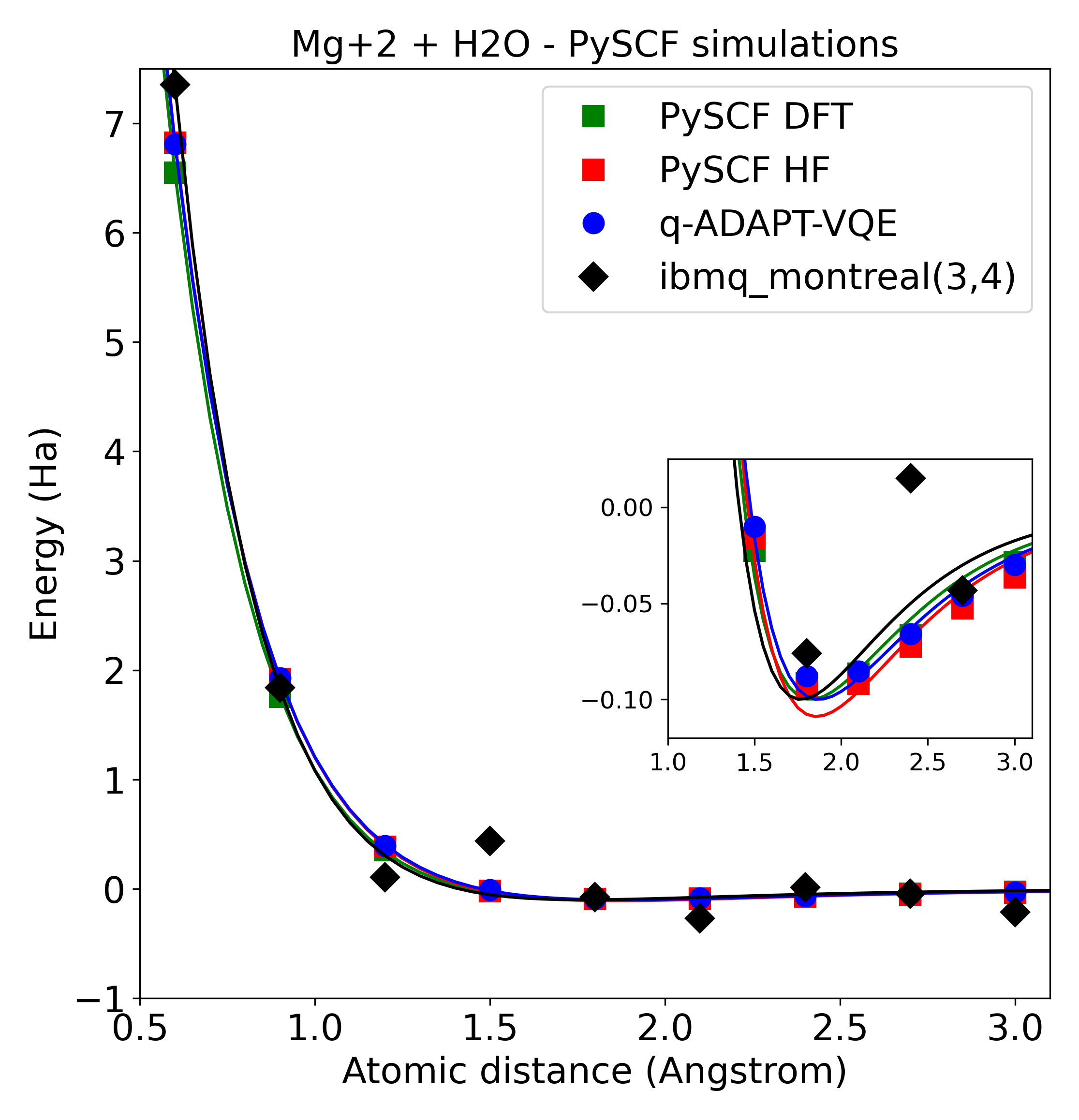}
    \caption{Mg\textsuperscript{+2} + H\textsubscript{2}O}
    \label{fig:Mg_H2O_pyscf_plot}
  \end{subfigure}
  \quad
  \begin{subfigure}[b]{0.3\textwidth}
    \centering
    \includegraphics[width=\textwidth]{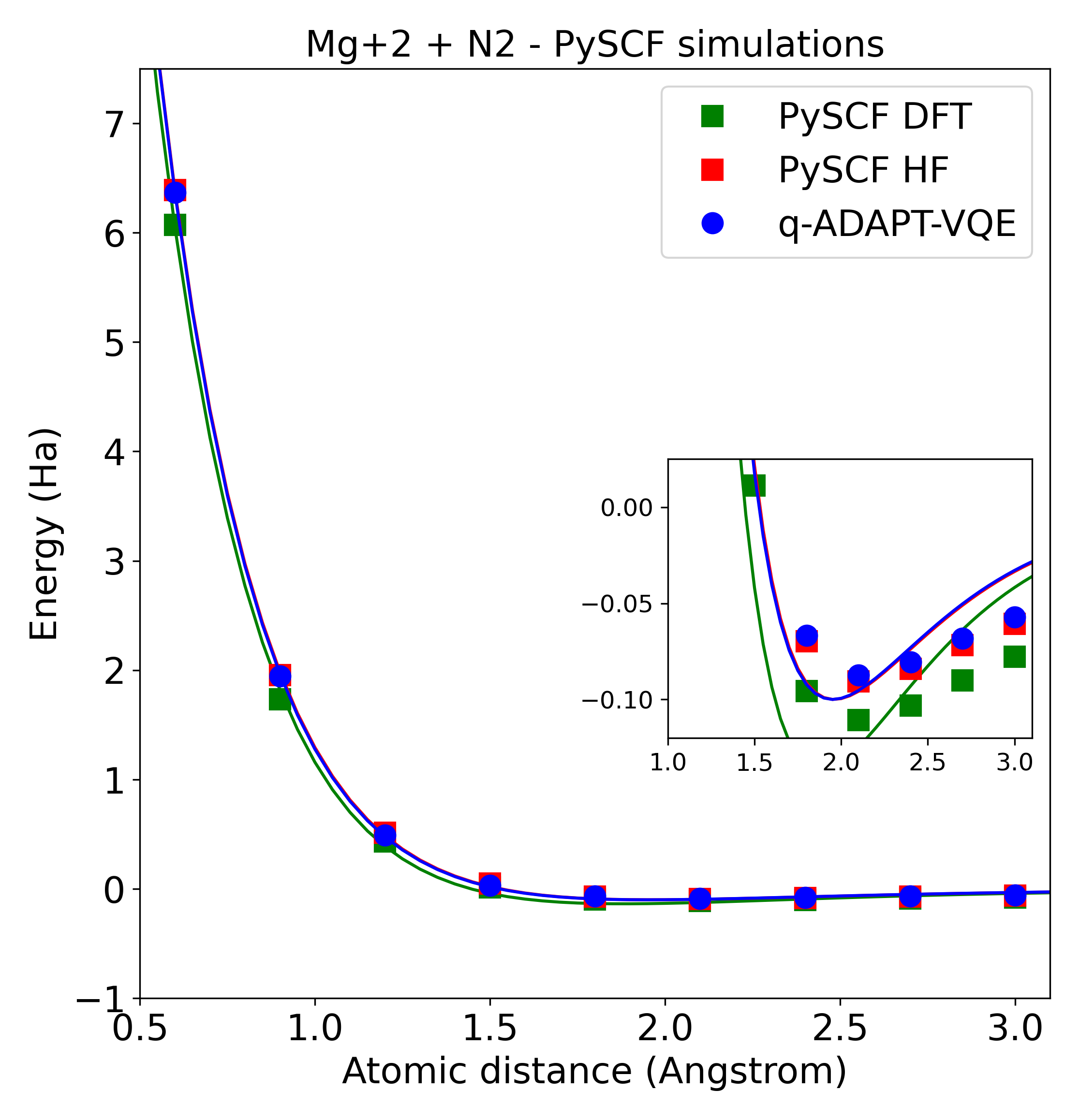}
    \caption{Mg\textsuperscript{+2} + N\textsubscript{2}}
    \label{fig:Mg_N2_pyscf_plot}
  \end{subfigure}
   \quad
  \begin{subfigure}[b]{0.3\textwidth}
    \centering
    \includegraphics[width=\textwidth]{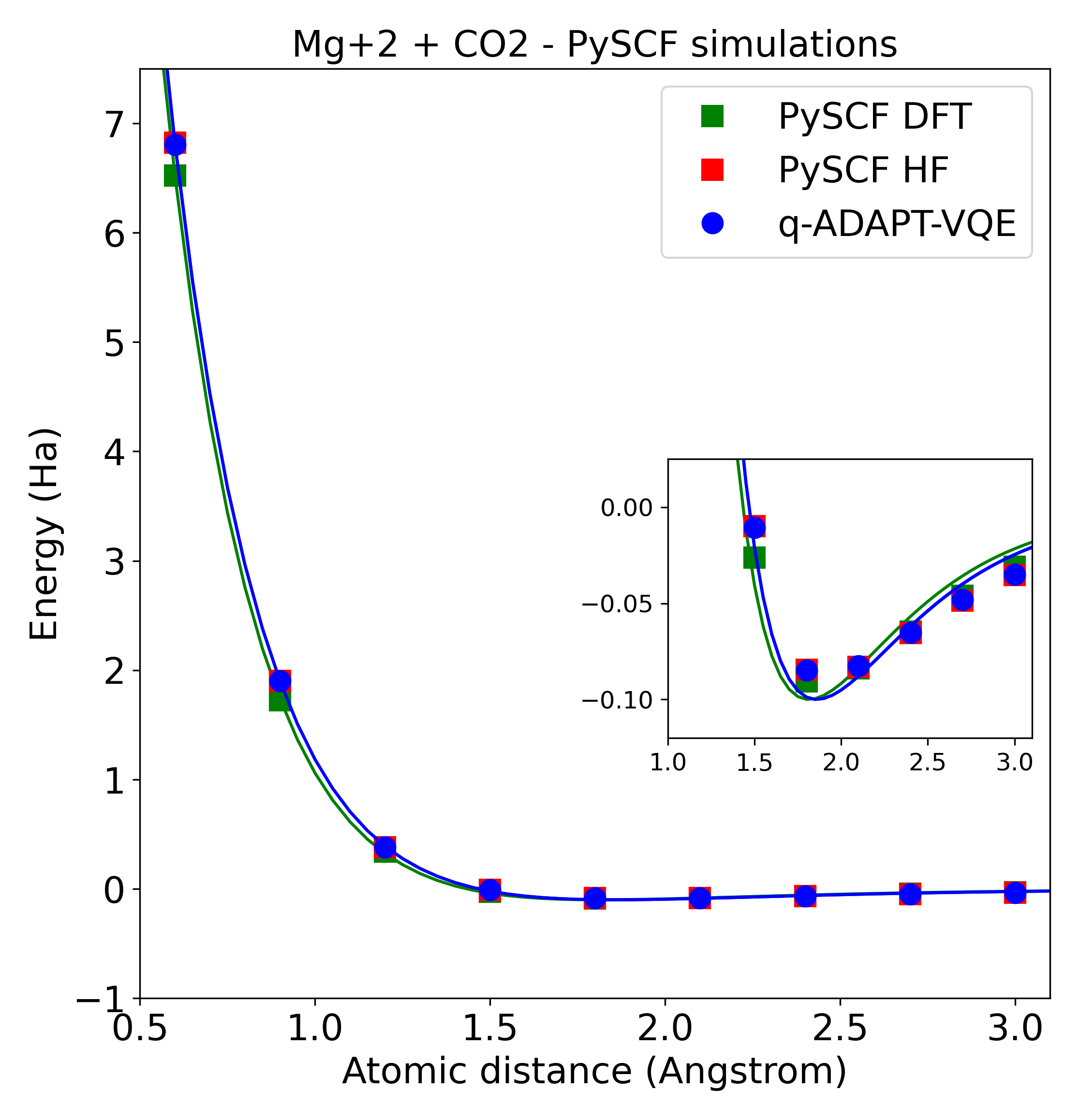}
    \caption{Mg\textsuperscript{+2} + CO\textsubscript{2}}
    \label{fig:Mg_CO2_pyscf_plot}
  \end{subfigure}
  \quad
\caption{Potential energy surfaces for the Mg\textsuperscript{+2} systems obtained from VQE experiments on quantum hardware using (3,4) active space (black diamonds) and VQE simulations in a (6,7) active space (blue circles) compared to PySCF-based classical benchmarks using the HF (red squares) and DFT (green squares) methods. The simulated qubit-ADAPT-VQE result matches the HF benchmark for the (a) H\textsubscript{2}O, (b) N\textsubscript{2} and (c) CO\textsubscript{2} molecules. The result from the hardware experiments, despite their inherent noise, also show accurate values within 0.1 Angstrom of both the simulated VQE and the classical benchmarks.}
\label{fig:Mg_pyscf_plots}
  
\end{figure}

In the following, we compare the potential energy surfaces computed using classical methods from PySCF, such as HF and DFT, to the best-in-class simulated VQE method, namely, qubit-ADAPT-VQE in a (6,7) active space. In Figure \ref{fig:Mg_pyscf_plots}, we show the agreement between quantum methods and their classical benchmarks. In particular, for Mg\textsuperscript{+2} + H\textsubscript{2}O, we also include in Figure \ref{fig:Mg_H2O_pyscf_plot} the comparison to the quantum hardware experiments running qubit-ADAPT-VQE with \texttt{ibmq\_montreal} in a (3,4) active space.

Overall, the simulated qubit-ADAPT-VQE results match the HF benchmark for all Mg\textsuperscript{+2}-based systems studied, with the DFT benchmark coming quite close. The PES obtained by running in the quantum hardware, despite its inherent noise, is also in good agreement with the classical benchmarks. This conclusion is supported not only by the visual inspection of Figure \ref{fig:Mg_pyscf_plots} but also by a comparison of the numeric results of the observables extracted from the PES fit in Table \ref{tbl:Mg_res_pyscf}. Here, the Binding Energy values all agree within the numeric uncertainty, and while the equilibrium distances between qubit-ADAPT-VQE and HF concur, they are found to be 2-3\% higher than what is predicted by DFT.

It is worth noting the high uncertainties in our hardware results that are due to application of error mitigation techniques. While accuracy increases we typically observe a precision loss. Considering the high uncertainties observed in our results, we decided not to use advanced error mitigation techniques that would increase those uncertainties even further.

\begin{table}
  \caption{Comparison of Equilibrium Distances and Binding Energies for Mg\textsuperscript{+2}-based systems obtained from the fitted Morse potential across different methods, both quantum (simulations and  hardware experiments) and classical. The Binding Energy values all agree to within the numeric uncertainty. The equilibrium distances agree between qubit-ADAPT-VQE and HF, but are 2-3\% higher than what is predicted by DFT.}
  \label{tbl:Mg_res_pyscf}
  \begin{tabular}{lllll}
  \hline
    \hline
        \multicolumn{5}{c}{Mg\textsuperscript{+2} + H\textsubscript{2}O}\\
    \hline
        & ibmq\_montreal & qubit-ADAPT-VQE & PySCF HF &PySCF DFT \\
    \hline
    Equilibrium Distance (Å)  & $1.8\pm0.3$ & $1.87\pm0.01$  &$1.868\pm 0.004$ & $1.83\pm0.02$  \\
    Binding energy (Ha)  & $-0.1\pm0.4 $& $-0.10\pm0.02$  &$-0.10\pm0.01$& $-0.1\pm0.02$  \\
    \hline
    \hline
        \multicolumn{5}{c}{Mg\textsuperscript{+2} + N\textsubscript{2}}\\
    \hline
        & ibmq\_montreal & qubit-ADAPT-VQE & PySCF HF &PySCF DFT \\
    \hline
    Equilibrium Distance (Å)  & - & $1.95\pm0.04$  &$1.96\pm 0.05$ & $1.90\pm0.07$  \\
    Binding energy (Ha)  & - & $-0.10\pm0.04$  &$-0.10\pm0.06$& $-0.1\pm0.08$  \\
    \hline
    \hline
        \multicolumn{5}{c}{Mg\textsuperscript{+2} + CO\textsubscript{2}}\\
    \hline
        & ibmq\_montreal & qubit-ADAPT-VQE & PySCF HF &PySCF DFT \\
    \hline
    Equilibrium Distance (Å)  & - & $1.86\pm0.02$  &$1.86\pm 0.02$ & $1.81\pm0.02$  \\
    Binding energy (Ha)  & - & $-0.10\pm0.02$  &$-0.10\pm0.02$& $-0.1\pm0.02$  \\
    \hline
  \end{tabular}
\end{table}

The experimental binding energies for CO\textsubscript{2} on Mg-MOF-74 are in the range of 39-47 kJ/mol.\cite{Demir2020} These values translate to 15-18 mHa per molecule, which is about 6x smaller than the results we obtained. This illustrates the limits of the oversimplification made by reducing the Mg-MOF-74 structure to a single metal ion. 

\newpage

\subsection{Conclusions}

As a key result of our investigation, the qubit-ADPAT-VQE achieves chemical accuracy when compared to UCCSD-VQE while maintaining hardware efficiency. This means it can give us an advantage when using quantum hardware by having a problem-specific ansatz that is also hardware efficient. Note that, despite the noisy results, we can fit the hardware data with a Morse potential so as to agree with Hartree-Fock predictions and the noiseless simulations.

Our results could be improved with more robust noise mitigation strategies, such as Zero Noise Extrapolation. It would also be of interest to explore larger systems and active spaces in order to improve our preliminary results. Although for that first we need to explore better techniques for active space selection.

Throughout our simulations, for Mg\textsuperscript{+2}
-based systems, we observed less satisfactory results when simulating the adsorption of N\textsubscript{2} molecules. Both the accuracy and the precision deteriorated across all methods used when compared to the other molecules. As a potential explanation, we speculate that the triple bond, which involves six electrons at once, is particularly hard to model. When the positively charged metal ion approaches, it perturbs at once all six active electrons involved in the triple bond. This seems to be especially problematic for the calculations limited to an active space, since the (6,7) active space is entirely consumed by these six electrons, leaving many other relevant electrons out of the calculation.

We also observed a systematic overestimation of the HF equilibrium distance with respect to the one obtained by DFT. This discrepancy was less than 5\% and may be attributed to the presence of exchange-correlation energy contributions that are absent in HF. This requires further investigation.


We note that current hardware limitations are mainly due to reliability, such as resistance to noise, and computation speed, meaning the number of circuits the hardware can run at a given time. The evolution of the available quantum volume and circuit-layer operations that can be performed at a given time interval will soon open pathways for larger-scale computation of CO\textsubscript{2} capture in MOFs for DAC applications.

\begin{acknowledgement}

The authors thank Max Rossmannek (IBM Research – Zurich) for providing help in implementing qubit-ADAPT-VQE and the broader Qiskit community for giving support. We thank Vidushi Sharma (IBM Research) for her support in the interpretation of the classical benchmark methods. We thank Felipe Lopes de Oliveira (IBM Research) for suggesting the Morse potential fit as a means to extract observables. We thank Ivan S. Oliveira (CBPF) and Alexandre Martins de Souza (CBPF) for insightful discussions on quantum computing and their support. The authors would like to acknowledge the support of Alexandre Pfeifer and Bruno Flach (IBM Research).

\end{acknowledgement}

\section{Code availability}

The code used to generate the results presented in this paper is available in a public GitHub repository.\cite{git_repo} This repository also includes additional experimental results that were not detailed in the manuscript. For ease of use, the repository contains a \texttt{README} file that explains how the additional results complement those discussed in the paper and guides the navigation of the supplementary material. Should there be any issues or questions regarding the code and its implementation, we encourage readers to reach out.

\bibliography{achemso-demo}

\end{document}